# SELF-EXCITED VIBRATIONS IN TURNING: EFFORTS TORSOR AND AVERAGE FRICTION COEFFICIENT ANALYSIS

Alain GERARD[1], Olivier CAHUC[1], Miron ZAPCIU[2]

**Abstract.** *An experimental device in turning including, in particular, a six-component dynamometer is exploited to measure the complete torque of cutting forces in a case of self-excited vibrations. For the tests, the tool used is type TNMA 16 04 12 carbide not covered (nuance carbide-SUMITIMO ELECTRIC), without chip breeze. The machined material is an alloy of chrome molybdenum type 42 CrMo24. The test-tubes are cylindrical with a diameter of 120 mm and a length of 30 mm. The effort of analysis relates to the moments. In particular, to the tool tip point, when feed rate increases the friction coefficient of swivelling is increasing while that of bearing is decreasing. Innovative remarks on the evolution of the average friction coefficient for various depths of cut and feed rate are presented.*

*Rezumat:* *In lucrare se utilizeaza o instalatie de achizitie de date experimentale in cazul strunjirii, in particular fiind utilizat un dinamometru cu sase componente, pentru masurarea fortelor si momentelor in cazul in care sunt prezente autovibratii in cadrul procesului. Pentru aceste teste scula aschietoare este de tip TNMA 16 04 12 din aliaj fara durificare superficiala (structura –SUMITIMO ELECTRIC), fara spargator de aschii. Materialul prelucrat este un aliaj cu crom si molibden de tip 42 CrMo24. Piesele de proba sunt cilindrice, cu diametrul de 120 mm si lungimea de 30 mm. Efortul de analiza se refera in mod special la momente. Sunt prezentate concluzii inovatoare relative la coeficientul mediu de frecare pentru diferite cazuri de variatie pentru adancimea de aschiere si avansul utilizat. Rezultatele prezentate corespund unui numar de aproximativ 7590 masuratori.*

## 1. Introduction

In the three-dimensional cutting case, the mechanical actions torsor (forces and moments) is often truncated: because the torsor moment part is probably a neglected fault of access to an adapted metrology [1-3]. Unfortunately, until now, the results on the cutting forces are almost still validated using platforms of forces (dynamometers) measuring those three components [4-5]. However, forces and pure moments (or torque) can be measured [6]. Recently, an application consisting in six component measurements of the actions torsor in cutting process was carried out for the case of high speed milling [7], drilling [8-9], etc. Cahuc et al., [10], present another use of this six-component dynamometer in an experimental study: the taking into account of the cut moments allows a better machine tool power consumption evaluation. This led to a better cut approach [8, 11, 12] and should enable us to reach new properties of the vibrations of the system piece-tool-matter.

---

[1]Prof., Dr, Université Bordeaux 1 (Sciences et Technologies) et CNRS Laboratoire de Mécanique Physique UMR 5496, Chair de Mécanique, 351 cours de la Libération, 33405 Talence, France, (e-mail: alain.gerard@u-bordeaux1.fr).
[2]Prof., Dr, Eng., Universitate Politehnica from Bucharest, 313 splaiul Independentei, MSP, 60042 Bucharest, Romania.

Moreover, the tool torsor has the advantage of being transportable in any space point and, in particular, at the tool tip in *O* point. This study follows the cut torsor and is carried out in several stages, including two major one. The first is related to the forces and the moments friction coefficient analysis at the tool tip during the cut. The second is dedicated to the torsor central axis determination during the cut. The central axes beams deduced from the multiple tests strongly confirm the moments presence to the tool tip. In a precise way, to section 2 we present primarily the experimental device used. Section 3 is devoted to measure the cut actions torsor reduced to the center of the six-component dynamometer and then transported to the tool tip point. An analysis of the forces exerted during the cut actions is carried out. It allows to establish in experiments certain properties of the resultant of the cut actions. The evolution of the average friction coefficient associate to a depth of cut for several feed rate is also obtained. The case of the moments to the tool tip point is examined. The torsor central axis is required (Section 4) and the central axes beams deduced from multiple measurements confirm the presence of moments to the tool tip point. In Section 5 we more particularly carry out the analysis of the moments to the central axis by looking at the case most sensitive to the vibrations ($ap = 5$ mm and $f = 0.1$ mm/rev). This study gives a certain number of properties. Before concluding, some innovative reflections are obtained, in the case of turning presented here.

## 2. Experimental device

The experimental device presented Figure 1 is a conventional lathe (Ernault HN400) for which the spindle speed does not exceed 3,500 rpm. The machining system behavior is analyzed through one three direction accelerometer fixed on the tool and using two unidirectional accelerometers positioned on the lathe, on the front bearing of the spindle, to identify the influence of this one on the cutting process. Moreover, a six-component dynamometer [6], being used as toolholder [13], is positioned on the lathe to measure all the cutting forces (forces and torques). On the test machining system, the instantaneous speed of rotation is controlled by a rotary encoder directly related to the workpiece. The connection is carried out by a rigid steel wire, which allows a better transmission of the behavior (Figure 1). During the cutting process, the number of revolutions is controlled constantly at nearly 690 rpm, and a negligible variation cutting speed of about 1% is detected. The cylindrical test-tubes have a diameter of 120 mm and a length of 30 mm (Figure 2). The dimensions retained for these test tubes were selected using the finite elements method coupled to an optimization method using SAMCEF® software presented in [14]. This process describes in [15] allows to ensure that the natural frequencies band of the machining system is well outer from that of the machine natural frequencies. Thus under a load P = 1,000 N, the material having a Young modulus E = $21.10^5$ N/mm² the workpiece dimensions selected are: $D_1$ diameter = 60 mm, $L_1$ length = 180 mm for a bending stiffness of
$7.10^7$ N/m (Figure 2). This value is including in the higher part of the interval of the acceptable rigidity values for conventional lathe [16-18].

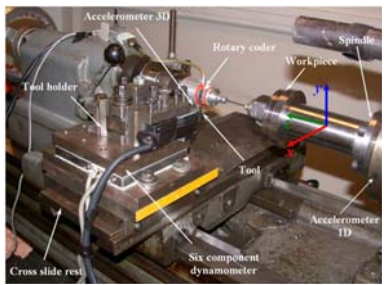
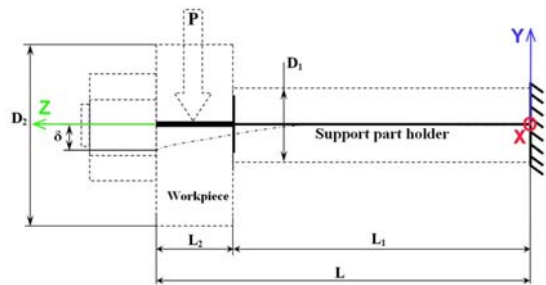

**Figure 1** Experimental machining system and metrology environment in turning process

**Figure 2** Geometry of holding fixture / workpiece

During the tests, the used tool was a noncoated carbide tool (TNMA 160412) without chip break geometry. The cutting material is a chrome molybdenum alloy type 42CrMo24. Moreover, the geometry of the tool (Figure 3) [19] is characterized by the cutting angle $\gamma$, the clearance angle $\alpha$, the edge angle of inclination $\lambda_s$, the direct angle $\kappa_r$, the nozzle radius $r_\varepsilon$ and the sharpness radius $R$. The tool insert is examined after each test and is changed if necessary of wear along the cutting face (Vb ≤ 0.2 mm ISO 3685), that may disturb the studied phenomenon. The tool characteristics used are presented in the Table 1.

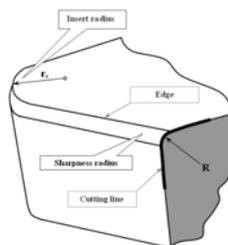

| $\gamma$ | $\alpha$ | $\lambda$ | $\kappa_r$ | $r_\varepsilon$ | R |
|---|---|---|---|---|---|
| -6° | 6° | -6° | 91° | 1,2 mm | 0,02 mm |

**Figure 3** The tool geometrical details    **Table 1** The tool geometrical characteristics

### 3. Cutting action torsor

#### 3.1. Tests

The experiments are performed within a framework similar to that exposed in Cahuc and al., [10]. For each test, the mechanical actions are measured according to the feed rate using the six-component dynamometer [7] following the method initiated in Toulouse [20], developed and finalized by Couétard [11]. These mechanical actions are evaluated for four depths of cut ap (= 1 mm; 2 mm; 3.5 mm; 5 mm) and according to four feed rate f (= 0.1 mm/rev; 0.075 mm/rev; 0.0625 mm/rev; 0.05 mm/rev).

Measurements are taken in the six-component dynamometer transducer O' center, and then transported to the tool point O via the moment transport traditional relations. Uncertainties of measurement with the six-component dynamometer used are about ± 4% for the forces components and ± 6% for the moments components. To check the repeatability and accuracy of identifications all tests and measurements are carried out four times and the average is selected for each configuration specified above. For each series of measure, we analyzed the results over 15 seconds of recording. We have 44 points of measurements recorded per full piece rotation. Thus, the average values discussed thereafter correspond finally to the average values of 7,590 measures.

#### 3.2. Resultant of the cutting forces analysis

Among the four values feed rate f indicated above, two examples of the cutting forces resultant measurements applied to the tool tip point are presented in the stable case (quasi non-existent vibrations) ap = 2 mm (Figure 4), and in the unstable case (self-excited vibrations) ap = 5 mm (Figure 5). In the stable case (ap = 2 mm, f = 0.1 mm/rev), we see that the amplitude of all forces components remain almost constant in the course of time. Only is detected a weak amplitudes variation about 1 to 2 N around their nominal values. Let us take for reference the absolute value of the component ($F_x$) of the forces. In the case of turning here studied, we observe that, whatever the depth of cut ap, there exists the following order relation between the modulus of the average cutting forces components:

$$|F_x| \leq |F_z| \leq |F_y|. \qquad (1)$$

The six-component dynamometer used gives the instantaneous values of all torque components of cutting action in the reference system of machine tool ($\vec{x}$, $\vec{y}$, $\vec{z}$) (Figure 1). The average values of the forces components ($F_x$, $F_y$, $F_z$) in this reference system are presented to Table 2 in the instability case (ap= 5 mm, feed rate variable f). To determine the average values of friction coefficients of the tool/piece it is initially necessary to have the forces components ($F_x$, $F_{ycop}$, $F_{zcop}$) in the tool reference system ($\vec{x}$, $\vec{y}_{cop}$, $\vec{z}_{cop}$). This is acquired by making the change of reference system [27]:

$$F_{ycop} = F_y \cos \alpha - F_z \sin \alpha; \quad F_{zcop} = F_y \sin \alpha + F_z \cos \alpha; \quad \alpha = 6° \text{ (Table 1)} \quad (2)$$

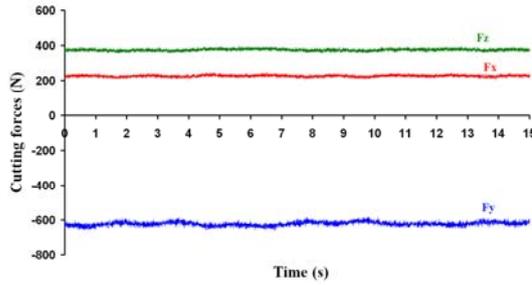 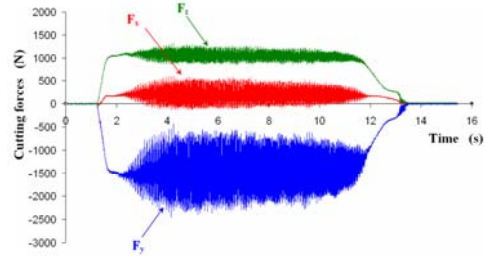

**Figure 4** Signals of the forces components exerted to the tool tip point according to the three directions of space machine in the stable case ap = 2 mm, f = 0.1 mm/rev, N = 690 rpm

**Figure 5** Signals of the forces components exerted to the tool tip point according to the three directions of space machine in the unstable case ap = 5 mm, f = 0.1 mm/rev, N = 690 rpm

In Eq. 2 the $F_{ycop}$ component is carried by the normal with the tool while the components $F_x$ and $F_{zcop}$ are in the plan of this one. Thus, it is proceeded to the determination of the average values tangential component modulus of the cutting forces in the tool plan $\|T_{cop}\|$. This one is given classically by the relation:

$$\|T_{cop}\| = \sqrt{F_x^2 + F_{zcop}^2}. \quad (3)$$

The average normal component modulus is: $\|N\| = \|F_{ycop}\|$. These elements allow to draw Table 2 the forces components average values applied to the tool in the reference system of machine tool, then in the reference system of the tool for a fixed depth of cut (ap = 5 mm) and a variable feed rate (f = 0.1 mm/rev; 0.075 mm/rev; 0.0625 mm/rev).

| f (mm/tr) | $F_x$ (N) | $F_y$ (N) | $F_z$ (N) | $F_{ycop}$ (N) | $F_{zcop}$ (N) | $\|T_{cop}\|$ (N) |
|---|---|---|---|---|---|---|
| 0,1 | 200 | -1512 | 1073 | -1616 | 909 | 946 |

Table 2. Forces components average values in the reference system of machine tool and tool for the unstable case ap = 5mm, f = 0.1 mm/rev and N = 690 rpm

These results lead to the average friction coefficient of Coulomb type represented in Figure 6. We see that the average values coefficient friction is a decreasing function of the feed rate. These results are coherent with those of the literature. However, in our case of three-dimensional cut we cannot neglect the $F_x$ component as in [28] for example.

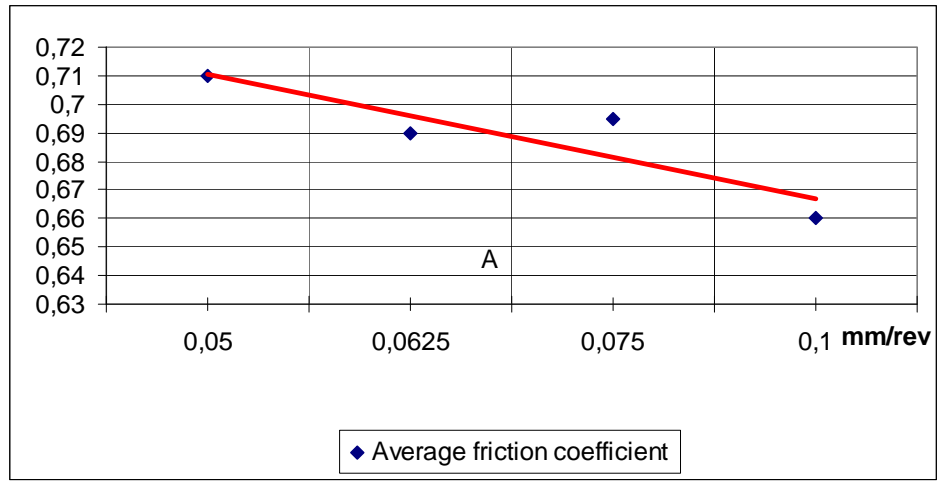

Figure 6. Average values friction coefficient evolution ($\|T_{cop}\| / \|N\|$) according to the feed rate for the unstable case (ap = 5 mm, f = 0.1 mm/rev and N = 690 rpm)

As for the torsor resultant, we give two examples of the moment components measurement to the tool tip point in the reference system of the machine tool. As comparison we present initially the stable case (without vibrations ap = 2 mm) Figure 7, then the unstable case (with vibrations ap = 5 mm) Figure 8.

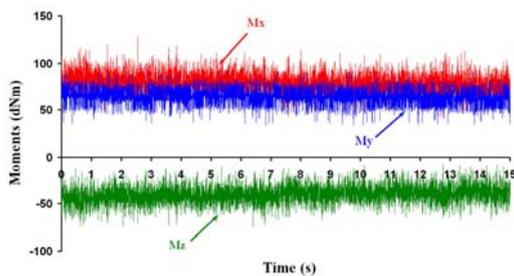 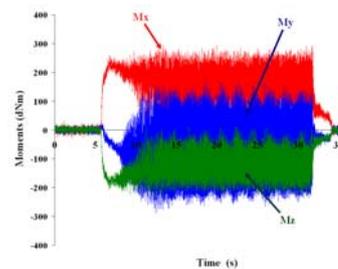

Figure 7. Signals of the moment components exerted to the tool tip point according to the three directions of the machine tool in the stable case ap = 2 mm, f = 0.1 mm/rev, N = 690 rpm

Figure 8. Signals of the moment components exerted to the tool tip point according to the three directions of the machine tool in the unstable case ap = 5 mm, f == 0,1 mm/rev, N = 690 rpm

As for the resultant in the stable case (ap = 2 mm, f = 0.1 mm/rev), the average moment components to the tool tip point are slightly disturbed while in the unstable case (ap = 5 mm, f = 0.1 mm/rev) the average values of the moment components of the tool tip point are definitely more chaotic.

However, one can note that the average moment components to the tool tip point are more disturbed than their equivalents on the level of the resultant. Thus, the moment components seem more sensitive to the self-sustained vibrations than the resultant components of the forces applied. The follow-up of the evolution of these could, for example, being a good means of detecting the existence of regenerative vibrations precociously.

As for the forces, on the moment components level we have, always in the case of turning examined here, the following order relation between the average values modulus of the moment components to the tool tip point (on 7,590 points of measurements).

$$|M_{oy}| \leq |M_{oz}| \leq |M_{ox}| \tag{4}$$

According to the confrontation of the Eq. (1) and the Eq. (4), it thus proves that on the order relations level between the modulus of the forces and the moment components average values to the tool tip point, the role of the $\vec{x}$ and $\vec{y}$ axes is reversed. This can be allotted to transport the moments from the tool tip point to the central axis.

In order to evaluate the average values friction coefficients of swivelling and bearing, we must now place ourselves in the reference system of the tool. Relations similar to Eq. (2) are then applied to the moment components transported to the tool tip point O. For the unstable case (ap =5 mm, f = 0.1 mm/rev) the moment average values are consigned to Table 3 where $||M_{otcop}||$ indicate the module of the average moment in the tool plan (cf. Eq. 3).

| f (mm/tr) | $M_{ox}$ (dNm) | $M_{oy}$ (dNm) | $M_{oz}$ (dNm) | $M_{ycop}$ (dNm) | $M_{zcop}$ (dNm) | $||M_{otcop}||$ (dNm) |
|---|---|---|---|---|---|---|
| 0,1 | 168 | -101 | -135 | -86,4 | -144,8 | 225,8 |

Tableau 3. Moment components average values of the forces in the reference system of the machine tool and this one of the tool for the unstable case ap = 5mm, f = 0.1 mm/rev and N = 690 rpm.

From this table of values, we deduce the average friction coefficients from bearing ($M_{o, Roul}$) and from swivelling ($M_{o, piv}$) to the tool tip point starting from the following traditional definitions:

$$||M_{otcop}|| = \sqrt{M_{ox}^2 + M_{ozcop}^2} \, , \, m_{o,Roul} = \frac{||M_{otcop}||}{||N||} \, ; \, m_{o,piv} = \frac{||M_{oycop}||}{||N||} \tag{5}$$

Figure 9 shows the evolution of these average friction coefficients according to feed rate. It is noted that the average friction coefficient of swivelling to the tool tip point is nearly constant even very slightly growing with f. On the other hand the average friction coefficient of bearing has a rather different behavior; this last is, all things considered, strongly decreasing when feed rate increases.

It is also noted that the average friction coefficients of bearing and swivelling are confused for the strong values feed rate (f = 0.075 mm/rev and f = 0.1 mm/rev). On the other hand for the low values feed rate (f= 0.05 mm/rev and f= 0.0625 mm/rev) these points are well dissociated. A thorough study relating to one more a large number of feed rate is planned in order to consolidate these results.

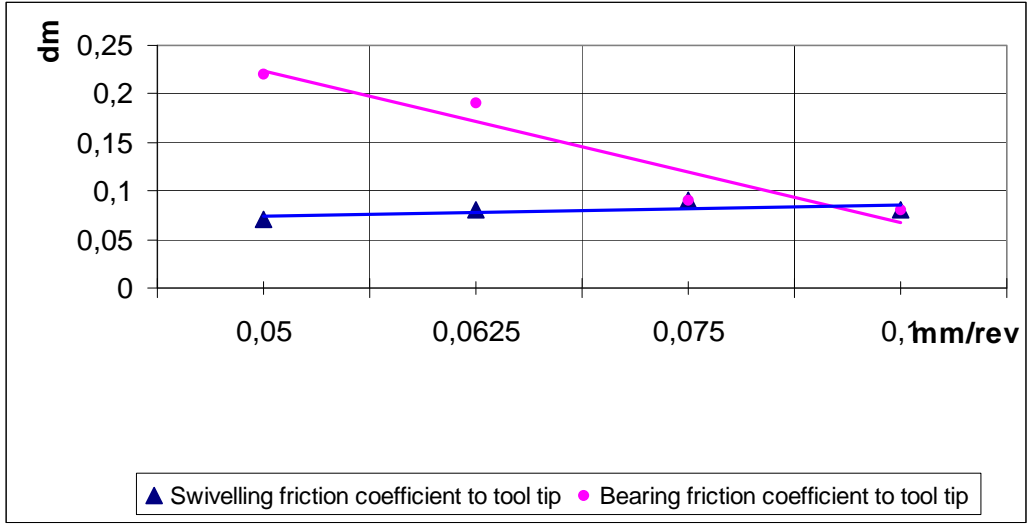

Figure 9. Swivelling and bearing friction coefficient evolution to the tool tip point.

## 4. Central axis determination

It is well-known that, with any torsor, it is possible to associate a central axis (except the torsor of pure moment), which is the single object calculated starting from the six components torsor [30]. A torsor $[A]_O$ in a point O is composed of a resultant forces $\vec{R}$ and the resulting moment $\vec{M}_o$:

$$[A]_o = \left\{ \begin{array}{c} \vec{R} \\ \vec{M}_o \end{array} \right. \tag{6}$$

The central axis is the line defined classically by:

$$\overrightarrow{OA} = \frac{\vec{R} \wedge \vec{M}_o}{\left|\vec{R}\right|^2} + \lambda \vec{R}. \tag{7}$$

In the Eq. (7) O is the point where the mechanical actions torsor was moved (here, the tool tip point ) and A is the current point describing the central axis.

Thus, OA is the vector associated with the bipoint [O, A] (Figure 10). This line (Figure 10a) corresponds to geometric points where the mechanical actions moment torsor is minimal. The central axis calculation consists in determining the points assembly (a line) where the torsor can be expressed according to a slide block (straight line direction) and the pur moment (or torque).

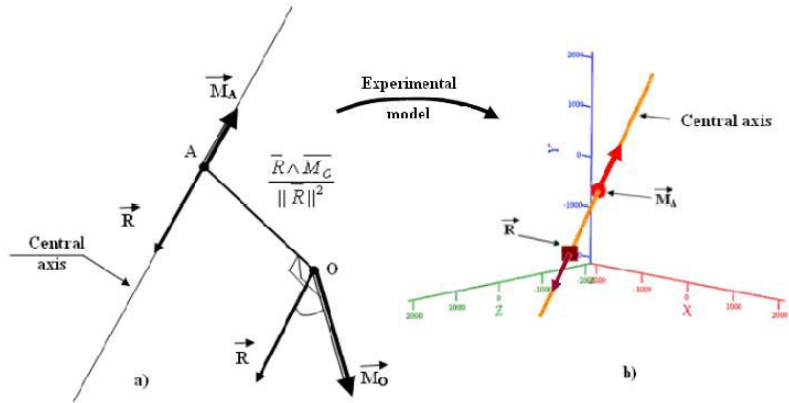

Figure 10. Central axis representation (a) and of the colinearity between vector sum $\vec{R}$ and minimum moment $\vec{M}_A$ on central axis (b).

The central axis is also the point where the resultant cutting force is colinear with the minimum mechanical moment (pure torque). The test results enable us to check for each point of measurement where the colinearity between the resultant cutting force $\vec{R}$ and moment $\vec{M}_A$ calculated is related to the central axis (Figure 10b). The meticulous examination of the six mechanical action torsor components shows that the forces and the moments average values are not null. For each measure point, the central axis is calculated, in the stable (Figure 11a) and unstable modes (Figure 11b). In any rigor, the case ap = 2 mm and f = 0.1 mm/rev should be described as "quasistable" movement because the vibrations exist but, as we already noticed, their amplitudes are very low, about the micrometer. Thus, they are quasi null compared to the other studied cases. Indeed, for example for ap = 5 mm and f = 0.0625 mm/rev, the recorded amplitude was 10 times more important.

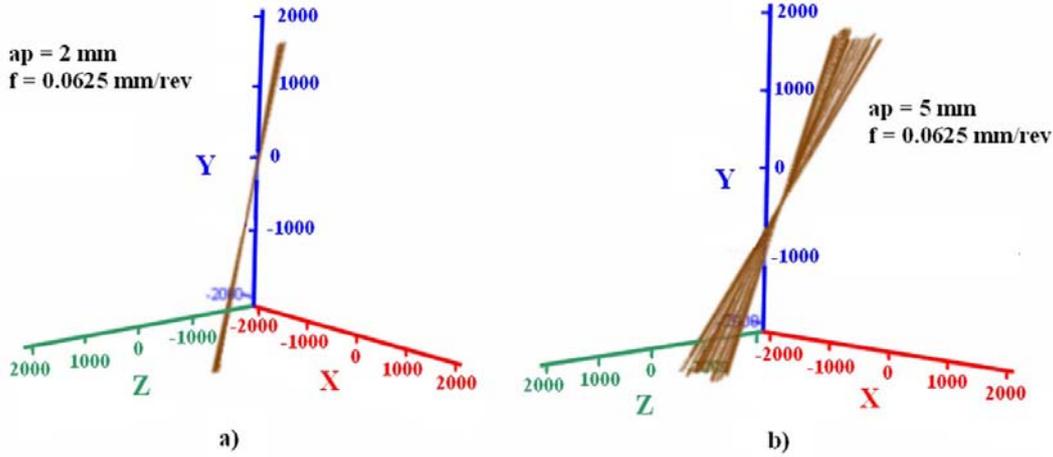

Figure 11 Central axes representation obtained for 68 rpm the workpiece speed and feed rate f = 0.0625 mm/rev (a) stable process ap = 2 mm; (b) unstable process ap = 5 mm

In the presence of vibrations (ap=5 mm), it is possible to observe for 68 revolutions of the machined piece the dispersive character of the central axes beam (fig. 11b), compared to the stable mode, where this same beam (fig. 11a) is tightened more or less tilted compared to the normal axis on the plan (x,y). This central axes dispersion can be explained by the self-sustained vibrations which cause the variable forces and moments generation.

## 5. Moments friction coefficient analysis at central axis

While transporting the moment from tool tip point to the central axis, the minimum moment (or pure torque) $\vec{M}_A$ is obtained. From the moment values to the central axis it is possible to deduce the constant and variable part of this one, as for the forces [24]. The analysis shows that the vibrations generate rotations of the tool, causes variations of contact and thus generates variable moments.

This representation allows to express the moments along the three axes of the machine tool: swivelling moment in the $\vec{y}$ direction and the two bearing moments following the $\vec{x}$ and the $\vec{z}$ direction, then in the ($\vec{x}$, $\vec{y}_{cop}$, $\vec{z}_{cop}$) reference system of the tool (Table 4).

| f (mm/tr) | $M_{Ax}$ (dNm) | $M_{Ay}$ (dNm) | $M_{Az}$ (dNm) | $M_{Aycop}$ (dNm) | $M_{Azcop}$ (dNm) | $\|M_{Atcop}\|$ (dNm) |
|---|---|---|---|---|---|---|
| 0,1 | -4,1 | -33,5 | 8,6 | -34,2 | 5 | 41,7 |

Table 4. Average values of the moment components to central axis in the machine tool reference system *(x, y, z)* and in the tool reference system ($\vec{x}$, $\vec{y}_{cop}$, $\vec{z}_{cop}$) for the unstable case (ap = 5 mm, f = 0.1 mm/rev and N = 690 rpm)

With the central axis the experimental data which preceding show, always in the case of turning studied here, the following order relation:

$$\|M_{Ax}\| \leq \|M_{Az}\| \leq \|M_{Ay}\|. \tag{8}$$

Thus, according to Eq. 1 the modulus of the average moments components to the central axis are in the same order as those of the average values force components. This is not the case to the tool tip point where the role of the axes $\vec{x}$ and $\vec{y}$ is reversed according to Eq. 4. Thus, the effect of the moment transport is sensitive.

Lastly, by using the central axis the definitions given in Eq. 5, we obtain the swivelling and the bearing average values coefficients of the pure torques. The evolution of those according to the feed rate is represented Figure 12. It is noted that the swivelling friction coefficient is constant to the central axis, while that of the bearing is slightly increasing when feed rate increases. This last tendency is opposite of that which we found to the tool tip, and in less accentuated. For the swivelling friction coefficient if it is constant on the central axis (Figure 12), on the other hand, to the tool tip (Figure 9) it slightly believes with feed rate (in the case of turning studied here). Thus one notes that with the central axis, the average values friction coefficients are distinct (with a common average value in swivelling and bearing for f = 0.075 min/rev). This is obviously a tangible consequence of more than influence of the moments transport and need for analyzing the pure torques in machining what is the principal originality of this work.

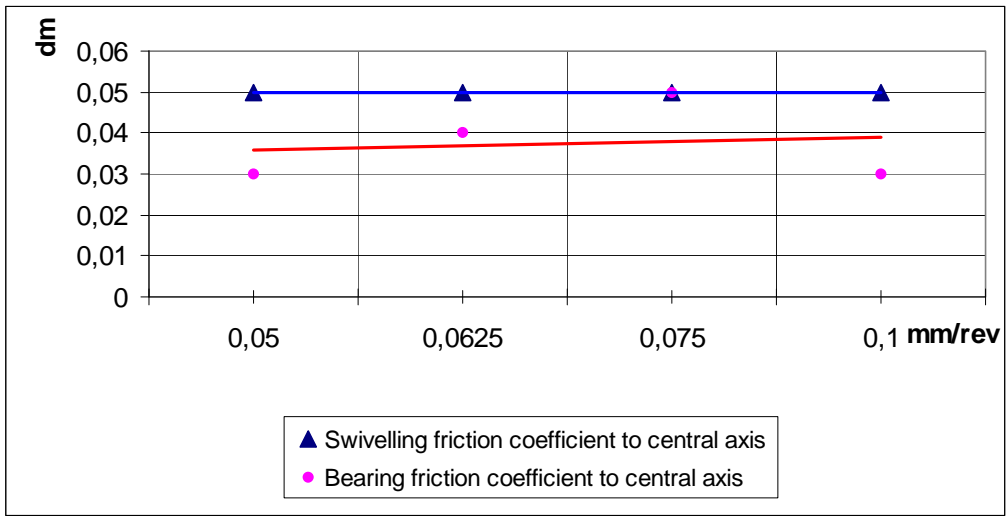

Figure 12. Friction coefficient evolution of swivelling and bearing moments to the central axis.

## 6. Conclusion

The experimental procedures installation allow to determine the elements necessary to a rigorous analysis of the average friction coefficients from the mechanical actions torsor. For the first time, it seems to us, is presented within the framework of the self-excited vibrations in turning a brief study of the resulting moments of the mechanical action torsor to the tool tip point and to the central axis. If the force components are compared with those of the moment, it is possible to note, that in absolute value, the average values are classified in the same order if the average values of the moment components are taken to the central axis. On the other hand, the roles of these same average values on the axes $\vec{x}$ and $\vec{y}$ are reversed if we look at the moments to the tool tip. The $\vec{z}$ axis however keeps a role of "pivot" (in the case of turning examined here).

Bearing and swivelling average values friction coefficients analysis allow to establish a rather clear difference between what occurs to tool tip and to central axis. In particular to the central axis, the average values friction coefficient of swivelling is constant what is not completely the case to the tool tip where a light growth is observed according to feed rate selected here. In the same way a notable difference exists between the average values friction coefficient of bearing to the central axis and to the tool tip; the evolutions are reversed. The decrease of the average friction coefficient of bearing is more important to the tool tip than the growth of this one to the central axis when the feed rate increases.

## R E F E R E N C E S